\documentclass[preprintnumbers]{revtex4}

\usepackage{graphicx}
\begin{document}
\bibliographystyle{revtex}

\preprint{IFT-01-35}

\title{Supersymmetric Lepton Flavour Violation}



\author{Jan Kalinowski}
\affiliation{Instytut Fizyki Teoretycznej, Uniwersytet Warszawski, 
Warsaw, Poland} 



\begin{abstract}
 In supersymmetric models new
sources of lepton flavour violation (LFV) may enhance the rates of charged
lepton flavour violating processes, like $\mu\to e\gamma$, and
generate distinct final states at future colliders, {\it e.g.} $\tau \mu+
jets + {E\!\!\!/}_T$.  We discuss the role of chargino pair production
for the LFV signals at $e^+e^-$ colliders.
\end{abstract}

\maketitle

\section{Introduction}
The neutrino experiment results \cite{P3-18neutos} have a natural
interpretation in terms of neutrino oscillations. This phenomenon is
violating the individual lepton flavour number, and raises an
interesting possibility of observing processes with charged-lepton
flavour violation, like $\mu\rightarrow e\gamma$,
$\tau\rightarrow\mu\gamma$, $\mu$-$e$ conversion in nuclei, rare $K$
decays etc.  Unfortunately, due to smallness of the neutrino 
masses, the predicted branching ratios for these processes are so tiny
that they are completely unobservable, e.g.  BR$(\mu \rightarrow e
\gamma) < 10^{-50}$.

In the
supersymmetric extension of the Standard Model, however, the situation
may be quite different. In addition to the SM mechanism, new
sources of lepton flavour violation (LFV) can be generated
by non-diagonal soft supersymmetry breaking slepton mass terms. 
Even if the off-diagonal slepton mass terms are assumed to vanish at tree 
level to avoid the supersymmetric flavour-changing problem, they
can be induced radiatively \cite{P3-18rad} in the framework of a seesaw
mechanism by the   right-handed heavy neutrinos.
In such a case a substantial $\nu_\mu - \nu_\tau$ mixing
leads to~\cite{P3-18feng,P3-18hisano} large $\tilde\mu_L - \tilde\tau_L$ and
$\tilde\nu_\mu - \tilde\nu_\tau$ mixings. 
As a result, the rates for rare processes, such as $\mu
\to e \gamma$, can be substantially enhanced.  
For nearly degenerate sleptons, however, these
contributions are suppressed through the superGIM mechanism by
$\Delta m_{\tilde{l}}/\bar{m}_{\tilde{l}}$ with the mass difference 
$\Delta m_{\tilde{l}}$ and the average mass $\bar{m}_{\tilde{l}}$ of
the sleptons.

Once superpartners are discovered, it will also be
possible to probe lepton flavour violation directly in their
production and decay processes \cite{P3-18kras}. 
 It has been
demonstrated that sneutrino or charged slepton pair production at
future $e^+e^-$ (and/or $\mu^+\mu^-$) colliders may provide a more
powerful tool to search for supersymmetric lepton flavour violation
(SLFV) than rare decay processes \cite{P3-18feng,P3-18nojiri}.
However, sneutrinos and charged sleptons may not
only be directly pair-produced in $e^+e^-$ collisions, but can also be
decay products of other supersymmetric particles, like charginos and
neutralinos, decaying via cascades.  We find that off-diagonal chargino
pair-production, overlooked earlier, can make a significant
contribution to the SLFV signal \cite{P3-18gkr}.


\section{Collider Signals of Supersymmetric Lepton Flavour Violation}

A flavor-violating signal is generated  by the production of 
real sleptons, followed by their oscillation into a different 
flavored slepton, and subsequent decay to a lepton.
For example, at $e^+e^-$ colliders events with
\begin{eqnarray}
\tau \mu+ 4j+ {E\!\!\!/}_T, \qquad\quad
\tau \mu l +2j+{E\!\!\!/}_T, \qquad\quad
\tau \mu l \bar l +2j+ {E\!\!\!/}_T \label{P3-18slfvsig}
\end{eqnarray}
can be expected.  Searching for such signals has several advantages:
first these processes are at tree level while rare decays are
generated by loop corrections. Second, the SLFV processes in decays of
sleptons are suppressed only as $\Delta
m_{\tilde{l}}/\Gamma_{\tilde{l}}$ \cite{P3-18ACFH205}, where
$\Gamma_{\tilde{l}}$ is the slepton decay width, as compared to the
$\Delta m_{\tilde{l}}/\bar{m}_{\tilde{l}}$ for rare decays.  Since
$\bar{m}_{\tilde{l}}/\Gamma_{\tilde{l}}$ is typically of the order
$10^2$--$10^3$, one may expect spectacular signals 
for possible discovery in future $e^+e^-$, $\mu^+ \mu^-$ or
$pp$ collider experiments. Last, but not least, the SM background with
two or more leptons with different flavours is quite small.

At $e^+e^-$ linear colliders the SLFV signals can be looked
for in decays of sleptons which are produced in pairs, 
\begin{eqnarray}
e^+e^- & \rightarrow &
\tilde{\ell}^-_i\tilde{\ell}^+_i \rightarrow  \tau^+\mu^- \tilde{\chi}^0_1
\tilde{\chi}^0_1, \nonumber \\
e^+e^- & \rightarrow &
\tilde{\nu}_i\tilde{\nu}^c_i  \rightarrow  \tau^+\mu^- \tilde{\chi}^+_1
\tilde{\chi}^-_1, \label{P3-18slelfv}
\end{eqnarray}
with $i=1,2,3$. These processes have been discussed in detail, showing
that they may be competitive to rare decay processes in 
searches for SLFV signals \cite{P3-18feng,P3-18nojiri}.

However, if the chargino $\tilde\chi^\pm_2$ is not much heavier, as is
the case in a substantial region of the MSSM parameter space, then
off-diagonal chargino or neutralino pair production $e^+e^-
\rightarrow \tilde\chi^\pm_1 \tilde\chi^\mp_2$, $\tilde\chi^0_1
\tilde\chi^0_2$ can take place for the linear collider CM energy
$\sqrt{s}=500 - 800 {\rm GeV}$. The heavier chargino and/or neutralino can
decay via the SLFV chain,
\begin{eqnarray}
e^+e^- & \rightarrow &
\tilde\chi^+_2\tilde{\chi}^-_1   \rightarrow  \tau^+\mu^- 
 \tilde\chi^+_1\tilde\chi^-_1,  \label{P3-18charlfv}\\
e^+e^- & \rightarrow &
\tilde\chi^0_2  \tilde{\chi}^0_1 \rightarrow  \tau^+\mu^-
\tilde\chi^0_1\tilde\chi^0_1, \label{P3-18neulfv}
\end{eqnarray}
where $\tilde{\chi}^\pm_1 \rightarrow \tilde{\chi}^0_1 f\bar{f}'$, and
$\tilde{\chi}^0_1$ escapes detection.  The signature therefore would
be the same as in slepton pair production, {\it i.e.}
eq.(\ref{P3-18slfvsig}),  
depending on hadronic or leptonic
$\tilde{\chi}^\pm_1$ decay mode.
As a result, the processes (\ref{P3-18charlfv}) and (\ref{P3-18neulfv}) provide
a new source for 
the signal in addition to those discussed in Ref.\cite{P3-18nojiri}.

\section{The role of charginos}

To illustrate the role of charginos for the SLFV process at an
$e^+e^-$ linear collider, such as the proposed TESLA~\cite{P3-18tdr}, 
we consider the signal and background
rates for two  representative points in the MSSM
parameter space. These points are given in terms of mSUGRA
scenarios defined by:
\begin{eqnarray}
&& RR1: m_0=100; 
\;\; M_{1/2}=200;\;\; A_0=0;\;\; \tan\beta=3;\;\;
{\rm sgn}(\mu)=+ \nonumber \\
&& RR2:  m_0=160; 
\;\; M_{1/2}=200;\;\; A_0=600;\;\; \tan\beta=30;\;\;
{\rm sgn}(\mu)=+
\label{P3-18eq:msugra}
\end{eqnarray}
chosen for detailed case studies at the ECFA/DESY
linear collider workshop. Here the masses and $A_0$ are in GeV, and
standard notation is used. 
The corresponding masses of chargino,
neutralino and slepton states relevant for SLFV processes at $\sqrt{s}=500$
and 800 GeV, along with some branching ratios, can be found in
\cite{P3-18gkr}.

To be more specific, we  take a pure 2-3 intergeneration mixing
between $\tilde\nu_\mu$ and $\tilde\nu_\tau$, generated by a
near-maximal mixing angle $\theta_{23}$, and  ignore any mixings
with $\tilde\nu_e$. The
scalar neutrino mass matrix $m^2_{\tilde\nu}$, restricted to the 2-3
generation subspace, can be written in the fermion mass-diagonal basis
as
\begin{equation}
{m}^2_{\tilde\nu} = \left(\matrix{\cos\theta_{23} & -\sin\theta_{23} \cr
\sin\theta_{23} & \cos\theta_{23}}\right) 
\left(\matrix{{m}_{\tilde\nu_2} & 0 \cr 0 &
{m}_{\tilde\nu_3}}\right) 
\left(\matrix{\cos\theta_{23} & \sin\theta_{23} \cr
-\sin\theta_{23} & \cos\theta_{23}}\right),
\label{P3-18one}
\end{equation}
where ${m}_{\tilde\nu_2}$ and ${m}_{\tilde\nu_3}$ are the
physical masses 
of $\tilde\nu_2$ and $\tilde\nu_3$ respectively.  In the following we
take the mixing angle $\theta_{23}$ and $\Delta m_{23} =
|{m}_{\tilde\nu_2} - {m}_{\tilde\nu_3}|$ as free, independent
parameters. 

In the case of RR2 the lightest chargino decays only leptonically
(without SLFV it decays into $\chi^0_1\tau\nu_\tau$).  
As a result, the signature of
SLFV in $\tilde\chi^\pm_2\tilde\chi^\mp_1$ production 
would be one muon and three taus plus
missing energy. Such a signal might be extremely difficult to extract from
background with four taus or with two muons and two taus for realistic
tau identification efficiencies.

For the case RR1, however, the lightest
chargino has a large branching ratio for hadronic decays. 
Therefore in our analyses we consider only their hadronic decay modes.
Given the mass spectrum, the off-diagonal
$\tilde\chi_1^\pm \tilde\chi_2^\mp$ pair is the only possibility for
the SLFV signal in chargino production at $\sqrt{s}=500$ GeV.  
The slepton flavour violation can occur in the
heavier chargino $\tilde\chi_2^\mp$  cascade decay chain as shown\\[1mm] 
\hspace*{10mm}[S1] ~~
$e^+e^- \rightarrow \tilde{\chi}^\pm_2 \tilde{\chi}^\mp_1 $, ~~~~
$\tilde\chi^+_2   \rightarrow  \tau^+ \tilde\nu_{2,3}$,   ~~~~
$\tilde\nu_{2,3} \rightarrow \mu^- \tilde\chi^+_1$\\[1mm]  
\hspace*{10mm}[S2] ~~
$e^+e^- \rightarrow \tilde{\chi}^\pm_2 \tilde{\chi}^\mp_1 $, ~~~~
$\tilde\chi^+_2   \rightarrow  \mu^+  \tilde\nu_{2,3}$, ~~~~
$\tilde\nu_{2,3} \rightarrow \tau^- \tilde\chi^+_1$\\[1mm]
followed by $\tilde\chi^\pm_1 \rightarrow \tilde\chi^0_1 + q + \bar q'$.  
There is another sequence with the charges reversed.  
The other process for the same final state, which was discussed
in~\cite{P3-18nojiri}, is the following\\[1mm]
\hspace*{10mm}[S3]~~
$e^+e^- \rightarrow \tilde\nu_i \tilde\nu_i^c$, ~~~~
$\tilde\nu_i  \rightarrow  \tilde\chi_1^- \tau^+$, ~~~~
$\tilde\nu^c_i   \rightarrow \tilde\chi_1^+ \mu^-$\\[1mm]
where $i=2,3$.  In [S1] and [S2] the
slepton flavour violating decay occurs in two ways leading to the the
same final state so that eventually the signal rate gets doubled.  
The background may originate from the flavour-conserving SUSY 
processes\\[1mm]
\hspace*{10mm}[B1]~~
$e^+ e^- \rightarrow \tilde\chi^+_2 \tilde\chi^-_1$, ~~~~
$\tilde\chi^+_2   \rightarrow  \tau^+  \tilde\nu_\tau$, ~~~
$\tilde\nu_\tau \rightarrow \tau^-(\rightarrow \mu^-)
\tilde\chi^+_1$\\[1mm]
\hspace*{10mm}[B2]~~
$e^+ e^- \rightarrow \tilde\chi^+_2 \tilde\chi^-_1$, ~~~~
$\tilde\chi^+_2  \rightarrow   \tau^+ (\rightarrow \mu^+)
\tilde\nu_\tau$, ~~~
$\tilde\nu_\tau \rightarrow \tau^- \tilde\chi^+_1$\\[1mm]
\hspace*{10mm}[B3]~~
$e^+ e^- \rightarrow \tilde\nu_i \tilde\nu^c_i$, ~~~~
$\tilde\nu_i   \rightarrow   \tilde\chi_1^- \tau^+$, ~~~
$\tilde\nu^c_i   \rightarrow \tilde\chi_1^+ \tau^-
(\rightarrow \mu^-)$\\[1mm]
\hspace*{10mm}[B4]~~
$e^+ e^- \rightarrow \tilde\tau_2^+ \tilde\tau_2^-$, ~~~~
$\tilde\tau^+_2  \rightarrow  \tau^+ \tilde\chi_2^0$, ~~~
$\tilde\chi_2^0 \rightarrow \tilde\chi_1^0 \tau^+(\rightarrow jets)
\tau^-(\rightarrow \mu^-)$, ~~~
$\tilde\tau^-_2  \rightarrow  \nu_\tau \tilde\chi_1^-$\\[1mm]
in which the $\mu$ comes from $\tau$ decay after
the $\tau\tau X$ events are produced, and from 
an important SM background\\[1mm]
\hspace*{10mm}[B5]~~ $e^+ e^- \rightarrow t \bar t g$\\[1mm]
with semileptonic top decays if two quark jets are allowed to overlap. 
At 800 GeV additional signal and background channels open \cite{P3-18gkr}.

\begin{figure}[htbp]
\centerline{
\includegraphics[width=6cm]{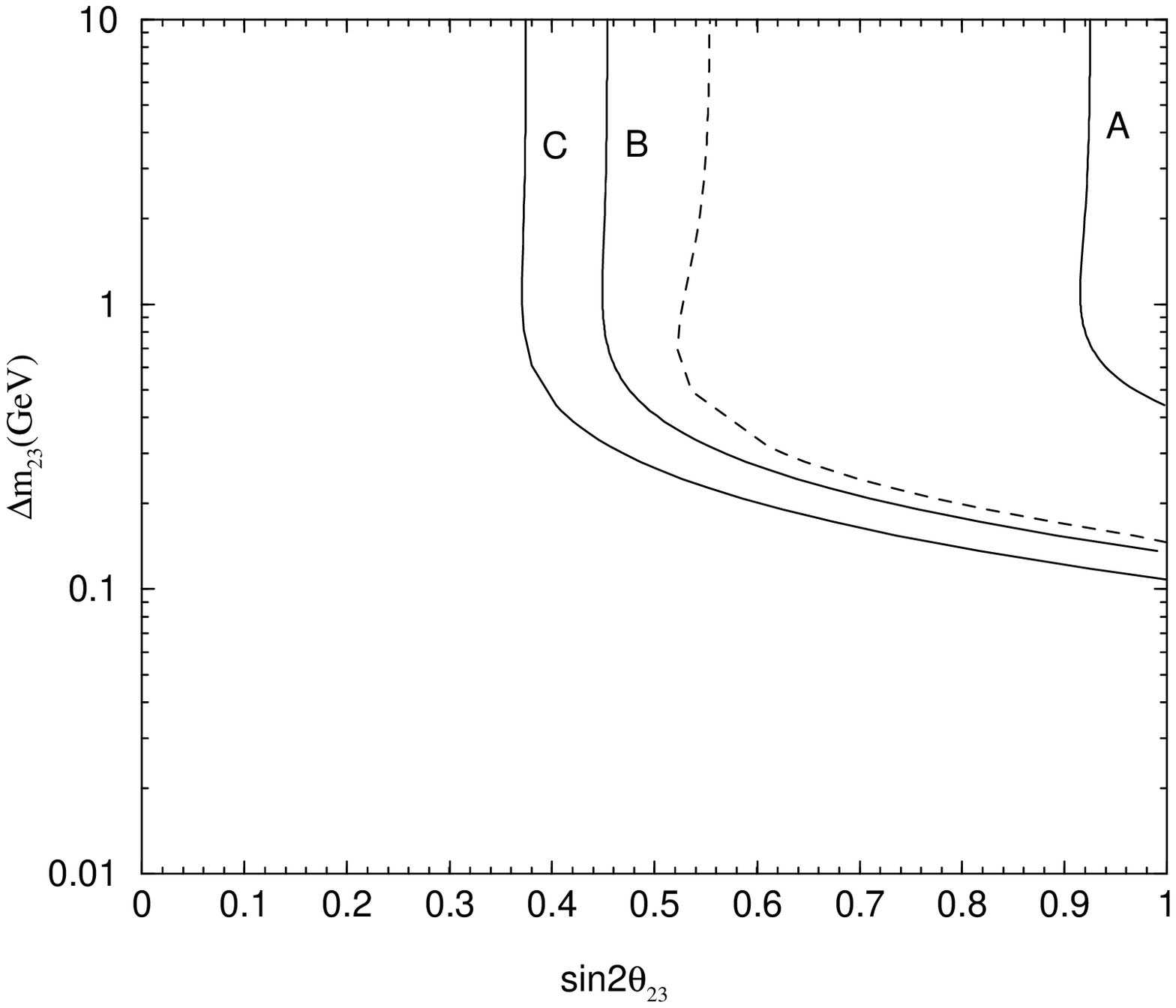} \hspace{1cm}
\includegraphics[width=6cm]{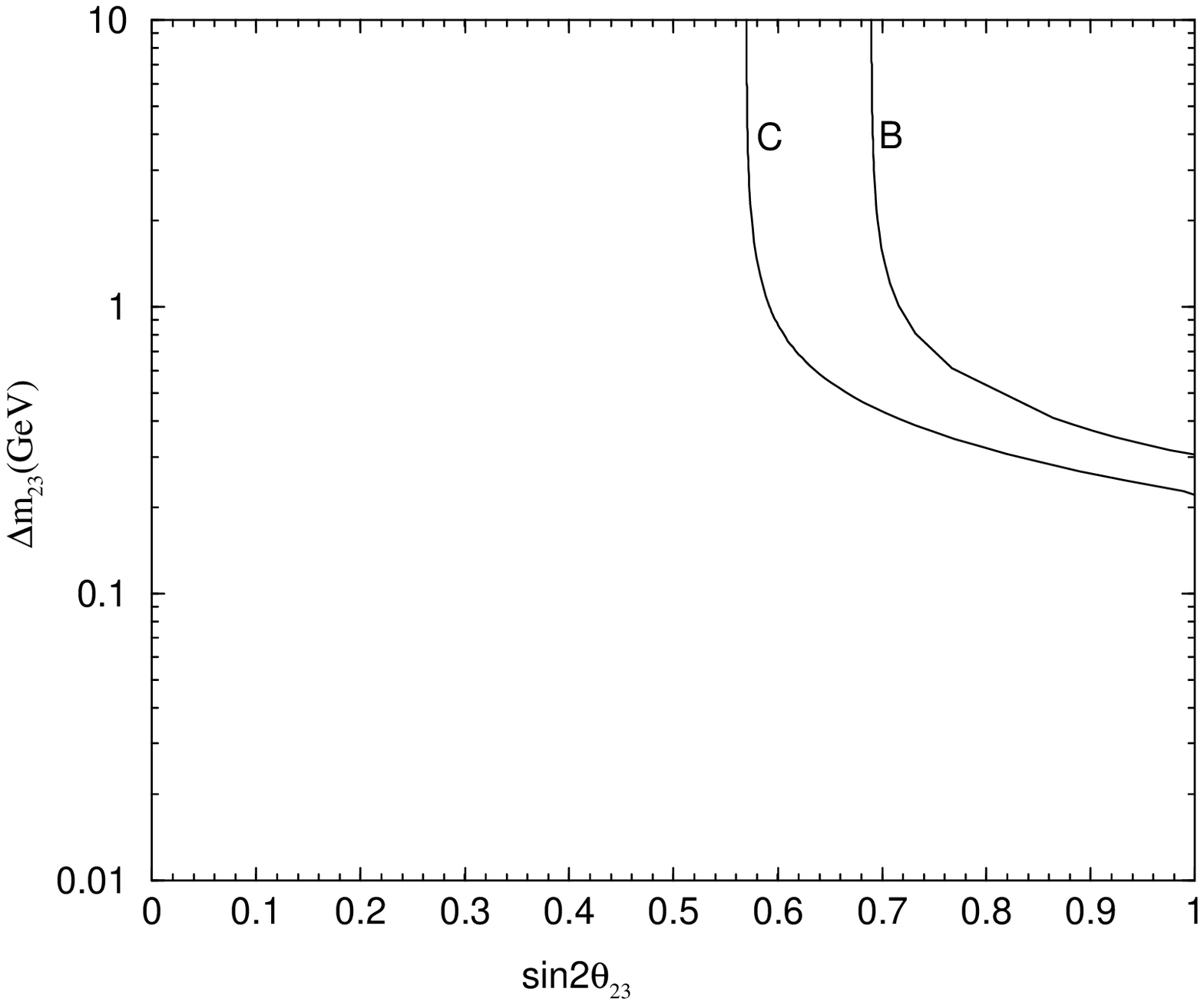}}
\vspace*{0.1cm}
\caption[]{{\it left: The significance contours for the SUSY point RR1 
  in $\Delta m_{23} -\sin2\theta_{23}$ plane for
 $\sqrt{s}=$500 GeV and for different luminosity options, contours A,
 B and C being for 50 fb$^{-1}$, 500 fb$^{-1}$ and 1000 fb$^{-1}$,
 respectively. The dashed line is for only $\tilde\nu \tilde\nu^c$
 contribution with luminosity 500 fb$^{-1}$.  The upper-right side of
 these contours can be explored or ruled out at the 3$\sigma$ level.}   \\
 {\it right: The  same for  $\sqrt{s}=$800 GeV and two
luminosities   
 500 fb$^{-1}$ and 1000 fb$^{-1}$ }}
\label{P3-18fig}
\end{figure}

With the set of kinematical cuts to ehnance the signal processes
(listed in \cite{P3-18gkr}), in Fig.\ref{P3-18fig}   the contour lines 
are plotted for the significance  
$\sigma_d =
\frac{N}{\sqrt{N+B}}$, where N and B is the number of signal and
background events respectively for a given luminosity. 
Comparing the dashed line with line 
B in Fig.1 we see that  that the chargino contribution [S1] and [S2]
increases the sensitivity range to $\sin^2\theta_{23}$ by 10-20\% while 
the sensitivity to $\Delta m_{{23}}$ does not change appreciably.
At 800 GeV the background is more important reducing the sensitivity
to the SLFV signal.

\section{Conclusions}
Neutrino oscillations imply the violation of individual lepton flavour
numbers and raise an interesting possibility of observing processes
with a violation of lepton flavour between two charged leptons. In
supersymmetry LFV processes may substantially be enhanced and many
interesting signals of SLFV processes may be expected at future
colliders. Our discussion of chargino contribution shows  
that a detailed account of all possible
production channels is needed in assessing the sensitivity of future
colliders to SLFV.

%
%

%
%


\begin{acknowledgments}
I am grateful to M. Guchait and P.Roy
for many stimulating discussions. The work was supported by the KBN
Grant  2 P03B 060 18  and 
the European Commission 5-th framework contract HPRN-CT-2000-00149. 
\end{acknowledgments}



\end{document}